# Light localization induced enhancement of third order nonlinearities in a GaAs photonic crystal waveguide

Alexandre Baron[1*], Aleksandr Ryasnyanskiy[1], Nicolas Dubreuil[1], Philippe Delaye[1], Quynh Vy Tran[2], Sylvain Combrié[2], Alfredo de Rossi[2], Robert Frey[1], Gerald Roosen[1]

[1]*Laboratoire Charles Fabry de l'Institut d'Optique, CNRS, Univ Paris-Sud, Campus Polytechnique, RD128, 91127, Palaiseau Cedex, France*
[2]*Thales Research and Technology, RD128, 91767, Palaiseau, France*
[*]*Corresponding author: alexandre.baron@institutoptique.fr*

**Abstract:** Nonlinear propagation experiments in GaAs photonic crystal waveguides (PCW) were performed, which exhibit a large enhancement of third order nonlinearities, due to light propagation in a slow mode regime, such as two-photon absorption (TPA), optical Kerr effect and refractive index changes due to free-carriers generated by TPA. A theoretical model has been established that shows a very good quantitative agreement with experimental data and demonstrates the important role that the group velocity plays. These observations give a strong insight into the use of PCWs for optical switching devices.

## 1. Introduction

Photonic crystal waveguides (PCWs) are known to exhibit peculiar propagation characteristics, such as strong transverse confinement and slow-light propagation. As a result, nonlinearities in such structures can be considerably enhanced. Therefore, PCWs are really interesting for developing very small and very fast optical devices. In that sense, it is important to master both the experimental and theoretical actions of slow-light induced light localization on nonlinearities. Nonlinear propagation measurements showing an enhancement of performances due to light localization at the band edge of 1D photonic crystals have previously been reported [1-3]. The effect of group velocity reduction has already been investigated in the case of electro-optic effects [4] and low command power optical switches have already been demonstrated in silicon [5], AlGaAs [6] and GaAs [7] microcavities. As a matter of fact it has very recently been demonstrated [8] that due to two-photon absorption (TPA) in GaAs cavities, nonlinearities can appear at a microwatt-level. So, combining slow-light properties of PCWs along with strong nonlinearities in III-V semiconductor devices such as GaAs, presents an undeniable opportunity, notably for optical switching.

The purpose of this letter is to report an experimental observation and a complete understanding of the third-order nonlinearities enhancement (TPA and self-phase modulation) due to reduced group velocity in a PCW. By doing so, generalization of the local field theory of nonlinear homogenous media to the case of PCWs is achieved. Self phase modulation in a AlGaAs PCW has already been observed [9], but the nonlinear behavior has not been related to group velocity reduction or local field effects. Our experiment is performed using thin GaAs photonic crystal membrane waveguides. In order to clearly address optical nonlinearities enhancement, we choose a sample in which group velocity reduction with low dispersion and long enough interaction length can be achieved. Even though low-group-velocities and low dispersion (LVLD) (with $v_g \simeq c/30$) have recently been achieved in PCWs [10-13], we decided to operate with the well-known W1 PCW for which fabrication and characterization are well mastered. In this structure a low dispersion value can be achieved in a moderate slow light regime ($v_g \simeq c/8$) [10] with a low disorder effect allowing to work with longer (≈1mm) PCWs. Our waveguides are 250 nm in thickness and the lattice period $a$ of the photonic crystal, consisting of air holes of $r = 0.3 \times a$ in radius, are 410 nm [14]. The guides are obtained by omitting one row of holes in the $\Gamma K$ direction and are 0.98 mm long.

## 2. Linear regime

Prior to nonlinear characterization, we achieved a linear transmission spectroscopy of the PCW, using a tunable CW laser around 1550 nm. Figure 1(a) shows the transmitted spectrum and reveals fringes due to the Fabry-Perot effect between the input and output facets of the guide. Using the decreasing spectral distance $\Delta \lambda$ between fringes, we are able to determine the group index $n_g = \lambda^2 / (2L\Delta\lambda)$ of the mode at different wavelengths. The measured $n_g$ are marked by crossmarks in Fig. 1(b). It is clear that $n_g$ increases with the wavelength and is largely superior to the bulk index $n_0 = 3.37$ in accordance with the classical dispersion band diagram for W1 waveguides [15]. A quadratic fit for $n_g$ is represented by a solid line and is used to calculate a group velocity dispersion (GVD) of $-7.10^5$ ps$^2$ km$^{-1}$ around 1550 nm. This reduction of the group velocity of the mode inside the waveguide is directly connected to light localization with a field enhancement in the waveguide that is customarily described using the local field factor $f$ [16,17]. In the case of slow modes, $f$ is equal to the square root of the slow down factor ($n_g/n_0$)[16,18] and is plotted in Fig. 1(b). As the third order nonlinearity is enhanced by $f^4$ [3,17], a large enhancement is expected for TPA and Kerr processes.

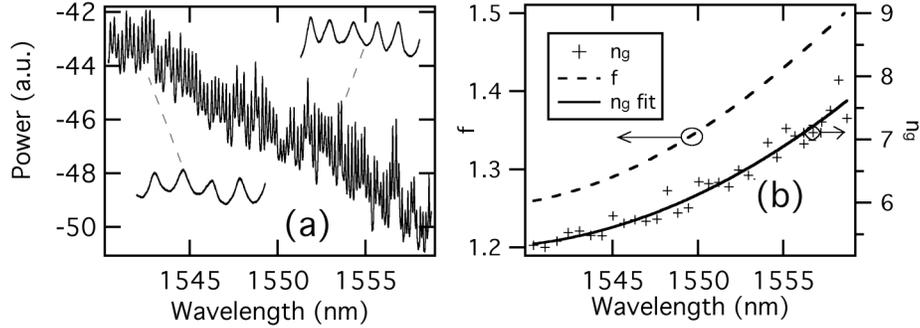

Fig. 1. (a) Linear transmission of the W1 guide (TE mode). Insets correspond to close-ups of the transmission for a 1 nm width. (b) Wavelength dependence of calculated group index (crossmarks). A quadratic fit is represented by the continuous line. The dashed line represents the f factor, calculated using the fitted curve of $n_g$.

## 3. Nonlinear regime

### 3.1 Experimental setup and results

Let us now emphasize on the nonlinear characteristics of the PCW in order to show the light localization induced enhancement of nonlinear interactions. To characterize our sample, around 1550 nm in the picosecond regime, we specifically developed an optical parametric oscillator (OPO) with a pulse duration of $\tau_L = 12$ ps and a repetition rate of $F = 80$ MHz. The pulses have a spectral width that is close to the Fourier limit (0.5 nm) [19], so as to have a fine characterization of the slow modes and resonant features of the PCW. Injection into the guide is achieved by using a microscope objective for which the numerical aperture is 0.85. Another objective, identical to the previous one, is used to collimate the transmitted beam at the output of the guide. The beam is then injected into a single-mode fiber that is linked to an optical spectrum analyzer (OSA) with a 0.02 nm spectral resolution.

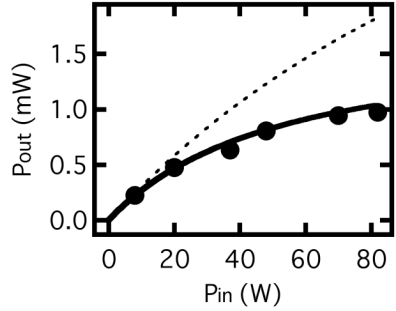

Fig. 2. Plot of power $P_{out}$ after output microscope objective vs power $P_{in}$ before input microscope objective. The thick and dashed lines denote the theoretical curve with and without slow light enhancement. Dots correspond to experimental data.

Figure 2 shows a power dependent transmission curve recorded at 1550 nm, where $n_g = 6.14$. $P_{in}$ and $P_{out}$ are the peak-powers at input and output of the injection and collection microscope objectives respectively and are obtained by multiplying the measured average power by $(F \times \tau_L)^{-1}$. Dots show the experimental data, which largely drifts away from the linear evolution demonstrating nonlinear absorption, due to TPA. Taking into account light localization in the PCW and neglecting linear absorption, which is estimated to be less than 1dB/mm for our guide [20], the output can be written as [21]:

$$P_{out} = K^2 P_{in} / (1 + K f^4 \beta_{TPA} L P_{in} / A_{eff}),  \qquad (1)$$

where $\beta_{TPA} = 10$ cm/GW [22] is the bulk TPA coefficient of GaAs, $A_{eff} = 0.1\,\mu m^2$ is the effective mode area and $L$ is the guide length. The coupling efficiency $K$ is assumed to be identical at input and output and was determined equal to $5.7 \times 10^{-3}$ from the linear transmission measurement that takes place at low power. Note that in equation (1), $\beta_{TPA}$ has been replaced by $f^4 \beta_{TPA}$, as it must be done for a third order nonlinear process [16] with $f = 1.35$, determined from Fig. 1(b). Using this correction, theory (continuous line on Fig. 2) and experiment (dots) are remarkably well in accordance with no adjustable parameters. If we neglect the local field correction, we obtain the dashed line, which is clearly unsatisfactory because it is significantly higher than the experimental results. These observations definitely demonstrate, as already noticed in [3], that the propagation of a beam with reduced group velocity cannot be described using the bulk material nonlinear coefficients. Note that the nonlinear absorption is 2.5 times larger in the PCW than in the bulk material which definitely demonstrates the major role played by group velocity reduction in the nonlinear interaction.

Second, we analyzed the transmitted spectrum (Fig. 3(a)) around 1554 nm, wavelength for which we observed the best results. The black and green line represent the spectra recorded at 8 W and 80 W peak power respectively. These powers were measured before the injection microscope objective. At lower power, the spectrum is identical to the input spectrum with the additional Fabry-Perot modulations observed in the CW regime (Fig. 1(a)). Notice that the pulse shape is expected to be conserved as the dispersive length $L_D = \tau_L^2 / GVD$ in our experiment (205 mm) is much larger than $L$. For an 80 W incident peak power, the spectrum has notably broadened. The Fabry-Perot spectral oscillations were eliminated by temporally filtering the signal autocorrelation function at half the round trip time inside the guide. As a matter of fact, by doing so, spectral interferences are suppressed because the first and second pulses are integrated without temporally overlapping. The filtered spectrum is represented by the red line on Fig. 3(a). After propagation inside the guide, the pulse undergoes both blue and red shifts, which are the result of self-phase modulation.

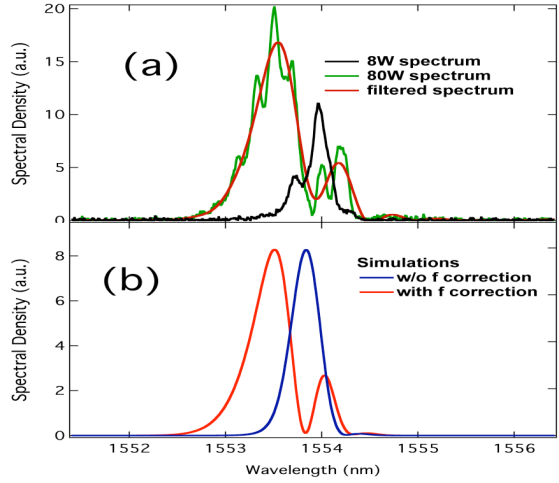

Fig. 3. (a) High and low injection power transmitted spectra through the guide (green and black curve respectively). The red curve represents the filtered high power spectrum. (b) Theoretical simulation of the transmitted spectrum for an 80 W incident peak power with (i.e. f=1.4) and without (i.e. f=1) slow light enhancement (red and blue line respectively).

We already have evidence of TPA as shown by the first experiment. Free-carriers induced by TPA have the effect of decreasing the index of refraction. If we hypothesize that carrier lifetime is much larger than the pulse duration, which is reasonable to assume given the dimensions of the guide [23], a refractive index change $\Delta n(z,t)$ proportional to the carrier density $N(z,t)$ is induced. Consequently, due to self phase modulation, the spectrum must exhibit a blue-shift, which indicates that free-carrier index change (FCI) does not fully explain the spectral split observed. This split is in fact due to self phase modulation induced by Kerr effect. Given the sign of the nonlinear Kerr index ($n_2 = 1.6 \times 10^{-17}$ m$^2$/W for bulk GaAs [22]), the leading edge of the pulse is red-shifted, whereas the trailing edge is blue-shifted. So Kerr effect and FCI have opposite signs on the leading edge, but have the same sign on the trailing edge, which means that a greater part of the pulse is blue-shifted. This is consistent with the observed spectrum, because the blue-shifted peak is stronger than the red-shifted one.

*3.2 Numerical simulation*

Beyond the qualitative agreement, we investigate the effect of the $f$ factor for a thorough quantitative description. In the same way $\beta_{TPA}$ is multiplied by $f^4$, $n_2$ should be multiplied by the same amount, because it is proportional to the real part of $\chi^{(3)}$. However, FCI is an effective fifth order process resulting from the mixing of a linear process (polarizability of free-carriers) and a third order nonlinear process (TPA generating free-carriers), for which the correcting local field factor is $f^6$.

Taking all the nonlinear processes into account, intensity $I(z,t)$ and phase $\varphi(z,t)$ of the propagating beam are governed by:

$$\partial I(z,t)/\partial z = -f^4 \beta_{TPA} I^2(z,t) \qquad (2)$$
$$\partial \varphi(z,t)/\partial z = k_0 [f^4 n_2 I(z,t) + f^2 \sigma_n N(z,t)], \qquad (3)$$

where $k_0$ is the free space wave vector and $\sigma_n = -7 \times 10^{-21}$ cm$^3$ is the refractive index change per carrier-pair density and can be calculated using the Drude model [23-25]. Integration of Eq. (2) leads to Eq. (1). Light intensity is governed by TPA (free-carrier absorption is neglected considering its low value [26]), while phase is influenced by optical Kerr effect and FCI due to TPA generated free-carriers with a density:

$$N(z,t) = \int_{-\infty}^{t} f^4 \beta_{TPA} (2\hbar\omega_0)^{-1} I^2(z,\tau) d\tau . \qquad (4)$$

Using Eqs. (2), (3) and (4), we numerically simulated the nonlinear propagation of the pulse through the guide. The result of this simulation enabled us to calculate the theoretical output spectrum for an 80 W incident peak power, which is plotted in red in Fig. 3(b). We observe a very good agreement between the simulated curve and the experimental curve from Fig. 3(a), considering that there are no adjustable parameters. The peak intensity of the injected pulse is determined using the measured value of $K = 6 \times 10^{-3}$, measured at low power for this experiment. The large nonlinear modification of the spectrum is therefore obtained for an injected peak power of only 0.5 W ($\simeq K \times 80$ W). Note that the simulated temporal phase reveals a maximum phase shift of almost $2\pi$ for the red curve on Fig. 3(b). Finally, for the same peak power, the simulated spectrum, obtained when neglecting light localization ($f = 1$), is represented on Fig. 3(b) in blue. It does not exhibit any clear spectral shift or broadening and clearly demonstrates the essential role of group velocity reduction in the enhancement of nonlinear effect at low peak powers.

## 4. Conclusion

In summary, we have performed nonlinear propagation experiments in a slow-mode PCW, demonstrating a large enhancement of nonlinear optical processes such as TPA, optical Kerr effect and refractive index changes due to TPA generated free-carriers. We clearly see that the refractive index change due to TPA-induced free-carriers, which is an effective fifth order process, is enhanced by a factor $f^6 \approx 8$ compared to the bulk material, while the Kerr nonlinear index is enhanced by $f^4 \approx 4$. The very good quantitative agreement between theory and experiment proves the soundness of our model and clearly shows that reduction of the group velocity can be exploited to optimize nonlinearities. Despite the moderate reduction of the group velocity of $v_g \approx c/7$ achieved in our waveguide, a maximum nonlinear phase shift of $\pi$ is predicted with a peak power of only 0.3 W. This performance is due to the strong nonlinear properties of GaAs and to our manufacturing control that allows the realization of millimeter length samples with low disorder effects. Stronger reduction of $v_g$ as obtained recently in LVLD PCWs [8-10], would be interesting for nonlinear applications but should be handled with care. Indeed increasing the local field factor $f$ enhances the higher order nonlinearities, such as three-photon absorption or free-carrier index changes [2], more rapidly than the lower order two-photon absorption or Kerr effect, leading to necessary trade-offs in the conception of nonlinear devices.

## Acknowledgments

A. Ryasnyanskiy acknowledges the RTRA "Triangle de la physique" for financial support.